\documentclass{arxiv2}

\begin{document}

\catchline{}{}{}{}{}
%%%%%%%%%%%%%%%%%%%%%%%%%%%%%%%%%%%%%%%%%%%%%%%%%%%%%%%%%%%%%%%%%%%

\title{The mass structure of SU(3) baryon antidecuplet and pion-muon mass difference 
}

\author{T. A. Mir}

\address{Nuclear Research Laboratory, Bhabha Atomic Research Centre, \\
Zakura, Srinagar - 190 006, Jammu and Kashmir, India\\
taarik.mir@gmail.com}

\maketitle

%\pub{Received (Day Month Year)}{Revised (Day Month Year)}

\section{Abstract}
We show that the mass intervals among the pentaquark baryons $\Theta^{+}$, $N^{*}(1685)$ and $\Xi_{3/2}^{-}$ proposed to be the members of the SU(3) antidecuplet are exact integral multiples of the mass difference between a neutral pion and a muon i.e. 29.318 MeV, in agreement with the general tendency of the mass differences among elementary particles to be close integral/half integral multiple of this mass unit. Further our study reveals that the strength of the SU(3) breaking understood to give members of the same multiplet different masses takes values, discrete as integral multiple of the mass difference between a neutral pion and a muon, the first two massive elementary particles. 
\section{Keywords}

pentaquarks; SU(3) antidecuplet; mass quantization.
%\end{abstract}

\ccode{PACS Nos.: 12.10.Kt, 12.40.Yx, 13.40.Dk}

\section{Introduction}

The most fundamental to elementary particles is the mass they posses, yet the generation of their observed masses is arguably the least understood feature of the standard model. On one hand the lepton and the quark masses are the free parametres of the standard model.\cite{Fritzsch} On the other hand the calculation of the hadron masses at the level of experimental accuracy has not been achieved owing to the confinement of quarks, the building blocks of hadrons, which makes the determination of their precise masses difficult. Further,  the standard model fails to explain the enormous range of the elementary particle mass distribution from a fraction of eV for neutrino to hundreds of GeV for the top quark.\cite{Veltman}
\newline
Since the recognition of any pattern amongst the elementary particle masses is a vital precursor in the development of standard model,\cite{Hall}$^{,}$\cite{Nambu} a systematic analysis of the experimetal mass data in search of any general relationships might provide an alternative approach to overcome the difficulities of this model. In fact the recognition of fascinating regularities in the spectrum of the hadrons resulted in the proposition of flavor SU(3) symmetry of strong interactions, the breaking of which is now known to give members of the same supermultiplet considerably different masses.\cite{Gell-Mann}$^{,}$\cite{Nee'man} The mass splitting within a given SU(3) multiplet are described in the Gell-Mann Okubo\cite{Okubo} mass formula leading to equal spacing rule for decuplet and antidecuplet.\cite{Oh} However, in the standard model no relationships are expected amongst lepton and hadron masses since leptons are described in QED, the theory of electromagnetic interaction, and hadrons are described in QCD, the theory of strong interactions. On the other hand in the description of both leptons and hadrons the associated physical mass is the most fundamental parameter, the existence of any general relationship amongst their masses will have direct consequences in understanding of their physics.  Thus, an analysis of the lifetime data has indicated alpha to be common scaling factor for both leptons and hadrons.\cite{Mac Gregor} The mass unit of 35 MeV equal to one-third of muon mass and one-fourth of the pion mass serves as a basic unit for the quantization of the both lepton and hadron masses.\cite{Nambu52,Palazzi} Similarly relationships amongst the masses of leptons have been reported.\cite{Barut}$^{,}$\cite{Koide}  Of late we have been reporting on the existence of another general relationship amongst leptons and hadrons that emerges through our specific mode of analysis. When arranged in the ascending order of their associated physical mass, the mass differences between the successive elementary particles turn to have a general tendency to be close integral/half integral multiple of mass difference between a neutral pion and a muon i.e. 29.318 MeV. The mass differences between unstable leptons and between baryons were shown to be quantized as integral multiples of this basic unit of 29.318.\cite{Shah} The effect was shown to be equally true for the neutral hadrons, octet and decuplet of baryons, pseudoscalar meson nonet and vector mesons.\cite{Mir} In the present study, we reveal that 29.318 MeV integral multiplicity of mass differences to be valid for mass differences of the SU(3) antidecuplet of pentaquark baryons. This shows that the mass difference between first two massive elementary particles i.e. a neutral pion  and a muon is a mass unit significant for the whole elementary particle mass system.

\section{Baryon Antidecuplet}
The prediction of an  antidecuplet consisting of pentaquark baryons in the chiral soliton model motivated the extensive experimental search to confirm the existence of these states.\cite{Dia} The lightest member of the antidecuplet $\Theta^{+}$ was discovered by LEPS Collaboration\cite{Nakano} and its existence was subsequently confirmed by other experimental groups.\cite{Barmin}$^{-}$\cite{Aslanyan} The $\Theta^{+}$ is an isosinglet state\cite{Jafe} with a mass of 1539.2 $\pm$ 1.6 MeV.\cite{Eidelman} The $\Xi_{3/2}^{-}$, member of the $\Xi$ multiplet of the antidecuplet was anounced by NA49 Collaboration and its mass is measured to be 1862 $\pm$ 2 MeV.\cite{Alt} $\Theta^{+}$ and $\Xi_{3/2}^{-}$ are the strange members of the antidecuplet. The existence and identification of the non-strange i.e. N members of the antidecuplet is under intense debate. The possible condidates for the N are the Roper resonance N(1440) and the N(1710).\cite{Dia,Jafe} In fact in the chiral soliton model, the mass the  of $\Theta^{+}$ was estimated as 1530 MeV in agreement with the experimentally observed value by identifying the N(1710) as the non-strange member of the antidecuplet.\cite{Dia} However, the association of N(1440) and the N(1710) with the antidecuplet has been challenged on many accounts.\cite{Glozman}$^{-}$\cite{Cohen} Quite recently the evidence for the existence of $N^{*}(1685)$, the non-strange member of the antidecuplet has been claimed.\cite{Kuznetsova} The mass of $N^{*}(1685)$ is 1685$\pm$5$\pm$7 MeV which is in the range 1650 - 1690 MeV, the theoritical  estimated value when the masses of the strange members $\Theta^{+}$ and $\Xi_{3/2}^{-}$ are taken as the input.\cite{Diakonov} The fourth member of the decuplet $\Sigma$ has not been detected. 
\newline
The splitting between the isomultiplets of the antidecuplet is one of the prime issues in any quark model.\cite{Stancu} Theoritically the mass differences between the members of the antidecuplet have been estimated on the basis of equal spacing rule derived along the lines similar for the decuplet from both chiral soliton model\cite{Dia} and SU(3) flavor symmetry.\cite{Oh} The equal spacing rule for the SU(3) decuplet and antidecuplet predicts masses of successive isospin multiplets to be equidistant. However, this rule is strictly obeyed niether for the decuplet nor for the antidecuplet since the mass separations are not exactly same. On the other hand we show that the mass splittings within the decuplet and antidecuplet members are exact integral multiple of the mass difference between a neutral pion and muon, an effect that appears to be valid for the whole elementary particle mass sysytem. In particular the average mass spacing among successive members of the baryon decuplet i.e. 146.816 MeV is very close to 146.59 MeV, a value obtained on integral (5) multiplication of 29.318 MeV. The difference between the observed and predicted values being 0.226 MeV only.\cite{Mir}   The effect we have been reporting is being argued to be a mere coincidence. On the other hand if the effect is real then one expects it to be valid for the antidecuplet also on the same lines as for other SU(3) multiplets.
\newline
The mass differences among the members of the baryon antidecuplet tabulated in Column 1 of Table 1 are given in Column 2.  Column 4 shows the integral multiples of 29.318 MeV that are close to the observed mass difference between members of the antidecuplet. The integers being shown in Column 3. The deviations of the observed value from the closest integral multiple of 29.318 MeV are given in Column 5. It is observed that the mass difference between $N^{*}(1685)$ and $\Theta^{+}$ i.e. 145.8 MeV differs from the nearest predicted value of 146.59 MeV by only 0.79 MeV. Same is true of the mass difference i.e. 177 MeV between the particles $\Xi_{3/2}^{-}$ and $N^{*}(1685)$ which differs from the predicted value of 175.908 MeV by only 1.092 MeV. 
Further, the observed mass difference between the heaviest and lightest members of the antidecuplet i.e. $\Xi_{3/2}^{-}$ and $\Theta^{+}$ i.e. 322.8 MeV differs from the integral (11) multiple of pion and muon mass difference by only 0.302 MeV. It may be noted that the deviation of the observed mass differences from those predicted as the integral multiple of 29.318 MeV fall within the uncertainity range of the experimentally measured masses of the $\Theta^{+}$, $N^{*}(1685)$, $\Xi_{3/2}^{-}$ and thus can be taken as zero. The fact that the mass differences of the leptons, mesons, baryons and pentaquark baryons are integral multiples of the mass difference between a neutral pion and  muon clearly indicates that the quark structure of the elementary particles seems to be irrelevent as for as 29.318 MeV integral multiplicity of their mass intervals is concerned. Further, our study reveals that the strength of the SU(3) breaking understood to give members of the same multiplet different masses can take only certain values, discrete as integral multiple of the mass difference between a neutral pion (hadron) and a muon (lepton).
\begin{table}[h]
\tbl{The observed mass intervals of baryon antidecuplet members as integral multiple of 29.318 MeV}
{\begin{tabular}{@{}lllllll@{}} \toprule
Particles & Mass Difference & Integer & $N$$\times$29.318  & Obsd - Expd  \\
& (MeV) & $N$  & (MeV) & (MeV)\\ \colrule
$N^{*}(1685)$ - $\Theta^{+}$\hphantom{00} & 145.8 & 5	& 146.59 & 0.79\\\\

$\Xi_{3/2}^{-}$ - $N^{*}(1685)$\hphantom{00} & 177 & 6	& 175.908 & 1.092\\\\

$\Xi_{3/2}^{-}$ - $\Theta^{+}$\hphantom{00} & 322.8 & 11 &  322.498 & 0.302 \\\\
\\
\botrule
\end{tabular} \label{ta1}}
\end{table}

\section{Conclusion}
The mass differences amongst the members of the baryon antidecuplet are found to be close integral multiples of the mass difference between a neutral pion and a muon. The fact that the mass differences among leptons, mesons, baryons and pentaquark baryons are integral/half integral multiple of mass difference between a neutral pion and a muon clearly demonstrates that 29.318 MeV multiplicity of elementary particles mass differences is independant of their structure.
%\section*{Acknowledgments}

\end{document}